\documentclass[conference]{IEEEtran}
\IEEEoverridecommandlockouts
\usepackage{hyperref}
\usepackage{titlesec}
\titlespacing{\section}{0pt}{10pt}{6pt}
\titlespacing{\subsection}{0pt}{8pt}{4pt}
\hypersetup{
    pdfborder={0 0 1},      % 开启边框，数字代表边框宽度
    citebordercolor={0 1 0} % 引用边框颜色设为绿色 (R G B: 0 1 0)
}
\usepackage{cite}
\usepackage{amsmath,amssymb,amsfonts}
\usepackage{algorithmic}
\usepackage{graphicx}
\usepackage{textcomp}
\usepackage{xcolor}

\graphicspath{{figs/}}

\def\BibTeX{{\rm B\kern-.05em{\sc i\kern-.025em b}\kern-.08em
    T\kern-.1667em\lower.7ex\hbox{E}\kern-.125emX}}

\begin{document}

\title{SPG-Codec: Exploring the Role and Boundaries of Semantic Priors in Ultra-Low-Bitrate Neural Speech Coding}

\author{
\IEEEauthorblockN{Mingyu Zhao$^{1,*}$, Zijian Lin$^{1,*}$, Kun Wei$^{2}$, Zhiyong Wu$^{1,\dagger}$\thanks{$^{*}$These authors contributed equally.}\thanks{$^{\dagger}$Corresponding author.}}
\IEEEauthorblockA{
$^{1}$Tsinghua Shenzhen International Graduate School, Tsinghua University, Shenzhen, China\\
$^{2}$Tencent, Shenzhen, China\\
\{zmy24, linzj24\}@mails.tsinghua.edu.cn, eldonwei@tencent.com, zywu@sz.tsinghua.edu.cn}
}

\maketitle

\begin{abstract}
Conventional neural speech codecs suffer from severe intelligibility degradation at ultra-low bitrates, where the bottleneck transitions from acoustic distortion to semantic loss. To address this issue, this paper conducts a systematic investigation into the role and fundamental limits of integrating frozen semantic priors – specifically HuBERT and Whisper – into neural speech coding. We introduce and quantitatively validate a novel Semantic Retirement phenomenon: while semantic constraints reduce the Word Error Rate (WER) by up to ${\sim}10\%$ relatively at 1.5 kbps, their benefits rapidly diminish beyond 6 kbps, indicating a practical capacity boundary. We further uncover a clear trade-off between different prior types: acoustic-rich priors (HuBERT) better preserve prosodic and timbral details, whereas high-level linguistic priors (Whisper) effectively suppress phonetic hallucinations in noisy environments (reducing hallucination rates by 26 percent) and substantially narrow the generalization gap for unseen speakers. Building on these findings, we propose a bitrate-aware regulation strategy that dynamically adjusts prior strength to optimize the trade-off between semantic consistency and perceptual naturalness. Extensive experimental evaluations confirm that our approach achieves competitive intelligibility and noise robustness compared to existing baselines, offering a principled pathway toward ultra-low-bitrate generative speech coding.
\end{abstract}


\section{Introduction}
\label{sec:intro}

The landscape of speech coding has undergone a paradigm shift with the advent of deep learning. While traditional digital signal processing (DSP) codecs like Opus and EVS have dominated telecommunications for decades, neural audio codecs have recently established new benchmarks for efficiency and quality. Early neural approaches explored WaveNet-based architectures and end-to-end optimization, laying the groundwork for modern systems. By integrating Residual Vector Quantization (RVQ) with Generative Adversarial Networks (GANs)~\cite{kong2020hifigan}, state-of-the-art architectures such as SoundStream~\cite{zeghidour2022soundstream}, EnCodec~\cite{defossez2023encodec}, and HiFi-Codec~\cite{yang2023hificodec} achieve high-fidelity reconstruction at bitrates as low as 3--6 kbps. Beyond compression, these discrete representations act as the fundamental vocabulary for the rapidly evolving field of Speech Large Language Models (SpeechLLMs)~\cite{wang2023valle,borsos2023audiolm,peng2025speechllm_understanding}, empowering systems like VALL-E~\cite{wang2023valle,chen2024valle2} and Moshi~\cite{defossez2024moshi} to generate human-parity speech.

Despite these advancements, a formidable barrier remains in the \textit{ultra-low bitrate} regime (e.g., $\le$ 1.5 kbps). In such extreme compression scenarios, the available bandwidth is often insufficient to encode both linguistic content and acoustic details (e.g., timbre, prosody) solely from the waveform. Consequently, traditional neural codecs suffer from severe degradation: phonemes collapse or blur, rendering speech intelligible but semantically ambiguous, a phenomenon we refer to as semantic collapse. Even recent large-scale trained models like BigCodec~\cite{xin2024bigcodec} and flow-matching approaches~\cite{chen2025f5tts} face challenges in balancing bitrate and semantic integrity. This indicates that the coding bottleneck has shifted from minimizing spectral distortion (an acoustic objective) to preserving linguistic integrity (a semantic objective), a challenge recently highlighted by~\cite{ye2024codecdoesmatterexploring}. To overcome this severe information bottleneck, the encoder must leverage external semantic priors to hallucinate or reconstruct plausible speech content from compressed cues.

In response to this challenge, a new wave of semantic-aware codecs has emerged. Innovations like SpeechTokenizer~\cite{zhang2024speechtokenizer} and RepCodec~\cite{huang2024repcodec} aim to disentangle semantic tokens from acoustic residuals. Similarly, frameworks such as SemantiCodec~\cite{liu2024semanticodec} and SNAC~\cite{siuzdak2024snac} employ multi-scale hierarchies to capture coarse-grained semantic features. Recent works like WavTokenizer~\cite{ji2024wavtokenizer} and DualCodec~\cite{li2025dualcodec} further align discrete codes with the semantic space of LLMs. While these methods demonstrate the potential of incorporating semantics, they primarily focus on architectural designs for generative tasks. A critical gap remains in understanding the fundamental mechanisms and limits of semantic guidance: \textit{(1) Under what conditions does semantic information become essential versus redundant? (2) How do different types of priors---ranging from acoustic-rich SSL features to abstract ASR transcriptions---impact the trade-off between intelligibility and naturalness?}

In this paper, we propose a unified analysis framework to systematically investigate the role of frozen semantic priors in ultra-low-bitrate speech coding. To this end, we augment a standard neural codec with auxiliary semantic conditioning from pre-trained foundation models, specifically HuBERT~\cite{hsu2021hubert} and Whisper~\cite{radford2023whisper}. This setup allows us to quantitatively isolate the benefits of semantic constraints and observe their interaction with bitrate allocation. Our contributions are fourfold:

\textbf{First}, we quantitatively define the \textbf{Semantic Retirement} phenomenon, identifying a 6 kbps boundary where priors shift from essential (reducing WER by ${\sim}10\%$ relatively at 1.5 kbps) to redundant. \textbf{Second}, we reveal a \textbf{trade-off} between prior types: HuBERT preserves acoustic texture, while Whisper suppresses hallucinations (by 26\%) and ensures linguistic consistency. \textbf{Third}, we demonstrate that semantic priors bridge the \textbf{generalization gap} on unseen speakers by over 15\% and enhance noise robustness. \textbf{Finally}, we propose a \textbf{bitrate-aware regulation} strategy that dynamically adjusts prior weights to balance semantic integrity and perceptual quality.

\section{Related Work}

\subsection{Neural Audio Codecs}
The transition from parametric coding to neural waveform modeling has revolutionized speech compression. Early neural approaches applied VQ-VAE~\cite{oord2017vqvae} to learn discrete representations from raw audio, though often suffering from spectral smoothing. The introduction of GAN-based discriminators~\cite{kong2020hifigan} significantly mitigated this by enforcing realistic phase and texture reconstruction. Building on these foundations, end-to-end codecs like SoundStream~\cite{zeghidour2022soundstream} and EnCodec~\cite{defossez2023encodec} employed Residual Vector Quantization (RVQ) to achieve scalable bitrates within a single model. Recent advancements continue to refine this paradigm: HiFi-Codec~\cite{yang2023hificodec} integrates Group-RVQ to reduce quantization errors, while DAC~\cite{kumar2023dac} addresses aliasing artifacts for high-fidelity music generation. More recently, large-scale models like BigCodec~\cite{xin2024bigcodec} and TiCodec~\cite{ren2024fewer} have pushed the boundaries of acoustic modeling. However, despite their success at medium bitrates (e.g., 6 kbps), purely acoustic-driven models face a theoretical bottleneck at ultra-low bitrates ($<2$ kbps), where the scarcity of bits leads to the collapse of phonetic structures, necessitating external information injection.

\subsection{Semantic-Aware Coding and Speech Generative Models}
To overcome the limitations of acoustic-only modeling, recent research has pivoted towards ``semantic-aware'' coding, largely driven by the needs of SpeechLLMs. One prevalent strategy is \textit{disentanglement}: SpeechTokenizer~\cite{zhang2024speechtokenizer} and RepCodec~\cite{huang2024repcodec} decouple speech into semantic tokens (for content) and acoustic tokens (for paralinguistics). Similarly, DualCodec~\cite{li2025dualcodec} and SACodec~\cite{dong2025sacodec} explicitly introduce semantic streams to enhance low-bitrate performance.
This trend aligns with the rise of generative audio models. Discrete token-based systems like VALL-E 2~\cite{chen2024valle2} and CosyVoice 3~\cite{du2025cosyvoice3} leverage these codes for zero-shot synthesis. Furthermore, the emergence of flow matching models, such as F5-TTS~\cite{chen2025f5tts} and E2 TTS~\cite{eskimez2024e2tts}, demonstrates that strong semantic conditioning can drive high-fidelity generation.
While these works successfully utilize semantic information for \textit{generation} or \textit{architectural design}, they lack a systematic, quantitative analysis of the \textit{boundary conditions}: specifically, at what exact bitrate does semantic guidance become redundant (``Semantic Retirement''), and how do different priors trade off between intelligibility and acoustic fidelity?

It is worth noting that our analysis framework is \textit{complementary} rather than competitive to these architectural works: the Semantic Retirement phenomenon and its 6 kbps boundary are diagnostic findings applicable across codec architectures, including SpeechTokenizer, DualCodec, and SemantiCodec.

\subsection{Self-Supervised Speech Representations}
Self-supervised learning (SSL) serves as the cornerstone for extracting semantic priors. Contrastive frameworks like wav2vec 2.0~\cite{baevski2020wav2vec2} and masked prediction models like HuBERT~\cite{hsu2021hubert} and WavLM~\cite{chen2022wavlm} capture rich acoustic and linguistic structures. WavLM specifically enhances robustness via denoising tasks, motivating our investigation into noise robustness. Conversely, weakly supervised models trained on massive labeled datasets, most notably Whisper~\cite{radford2023whisper}, demonstrate exceptional capabilities in ASR, capturing high-level semantics invariant to speaker characteristics.
In this paper, rather than training new features, we employ these pre-trained models as frozen priors to regulate the codec. By contrasting acoustic-heavy priors (HuBERT) with semantic-heavy priors (Whisper), we provide a granular analysis of their respective roles in bridging the bit-rate gap.
% --------------- METHOD START ---------------
\section{Methodology}

\begin{figure*}[t!]
\centering
\includegraphics[width=0.75\linewidth]{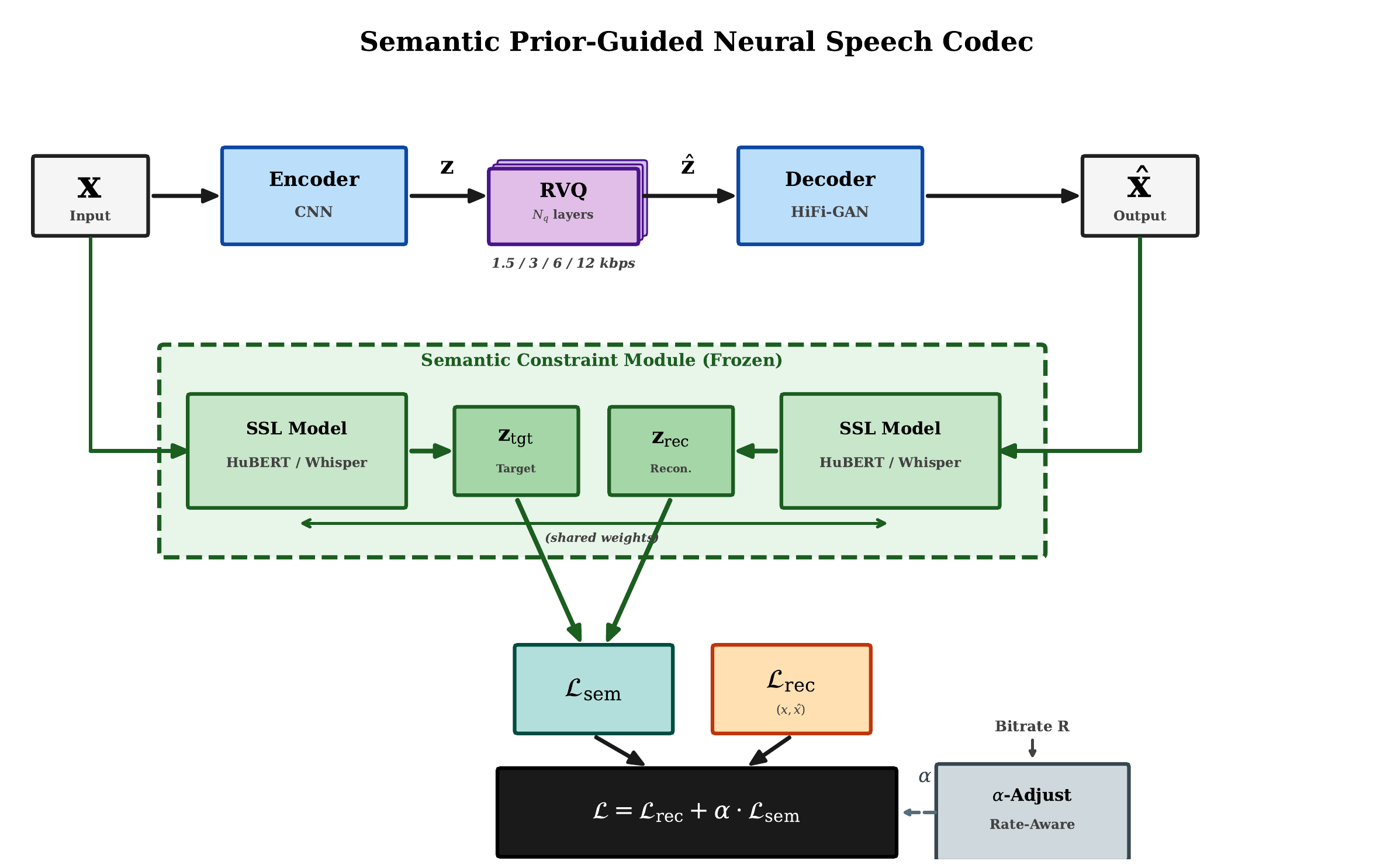}
\caption{Framework of our proposed SPG-Codec. A frozen Semantic Constraint Module (HuBERT/Whisper) guides the backbone, while the $\alpha$-Adjust module dynamically modulates semantic weight based on bitrate to address ``Semantic Retirement''. \textit{Shared weights}: a single frozen encoder instance is applied twice in forward-pass mode---once to $x$ to produce $z^\Phi_{\text{tgt}}$ and once to $\hat{x}$ to produce $z^\Phi_{\text{rec}}$---with no additional learnable parameters.}
\label{fig:framework}
\end{figure*}

In this section, we systematically analyze the interplay between acoustic fidelity and semantic guidance in ultra-low-bitrate coding. To facilitate this analysis, we propose a unified framework (as illustrated in Fig.~\ref{fig:framework}) comprising three core components: a scalable neural codec backbone, a frozen semantic constraint module, and a bitrate-aware regulation strategy. This design aims to quantitatively isolate the contribution of semantic priors and dynamically balance semantic consistency with perceptual quality.

% Figure 1: Framework

\subsection{Backbone Neural Codec}
\label{sec:backbone}
We adopt a convolutional encoder-decoder architecture akin to SoundStream~\cite{zeghidour2022soundstream} and EnCodec~\cite{defossez2023encodec} as the backbone.
Let $x \in \mathbb{R}^T$ be the input waveform. The encoder $E$ downsamples $x$ into a latent representation $z = E(x) \in \mathbb{R}^{D \times T'}$, capturing local acoustic features.

\textbf{Scalable Quantization.} To support variable bitrates (1.5--12 kbps) within a single model, we employ Residual Vector Quantization (RVQ). The quantizer $Q$ approximates $z$ as a sum of discrete vectors:
\begin{equation}
    \hat{z} = \sum_{k=1}^{N_q} \hat{z}_k, \quad \text{where } \hat{z}_k = \mathop{\arg\min}_{c \in \mathcal{C}_k} \|r_{k-1} - c\|_2
\end{equation}
Here, $r_0 = z$, $r_k = r_{k-1} - \hat{z}_k$ is the residual, and $\mathcal{C}_k$ is the codebook for the $k$-th layer. The number of active quantizers $N_q$ is dynamically adjusted during inference to determine the bitrate $R$.

\textbf{Generator \& Discriminator.} The decoder $G$ reconstructs the waveform $\hat{x} = G(\hat{z})$ using a HiFi-GAN-based architecture~\cite{kong2020hifigan}. To ensure perceptual quality, the training objective includes a multi-resolution spectral loss $\mathcal{L}_{spec}$ and an adversarial loss $\mathcal{L}_{adv}$ derived from a Multi-Period Discriminator (MPD) and a Multi-Scale Discriminator (MSD).

\subsection{Semantic Constraint Module}
Standard acoustic losses (e.g., $L_1$ and Spectral) penalize waveform differences but often fail to preserve phonemic integrity at ultra-low bitrates ($\le 1.5$ kbps), leading to ``mumbled'' speech. To mitigate this, we introduce a \textbf{Semantic Constraint Module} that injects linguistic priors into the reconstruction process. We investigate two distinct paradigms:

\textbf{1) Acoustic-Semantic Prior (HuBERT~\cite{hsu2021hubert}):} HuBERT is trained via masked prediction of hidden units. It effectively captures both phonetic content and paralinguistic features (e.g., prosody, speaker identity). We extract representations from the intermediate transformer layers, which are known to encode rich acoustic-phonetic information:
\begin{equation}
    z^{H}_{tgt} = \Phi_{\text{HuBERT}}(x), \quad z^{H}_{rec} = \Phi_{\text{HuBERT}}(\hat{x})
\end{equation}
Here, $\hat{x}$ is resampled if necessary to match the input sampling rate of the prior model (16 kHz).

\textbf{2) Linguistic-Semantic Prior (Whisper~\cite{radford2023whisper}):} Trained on 680k hours of weak supervision, Whisper's encoder representations are highly invariant to noise and speaker characteristics, focusing purely on linguistic content.
\begin{equation}
    z^{W}_{tgt} = \Phi_{\text{Whisper}}(x), \quad z^{W}_{rec} = \Phi_{\text{Whisper}}(\hat{x})
\end{equation}
We hypothesize that Whisper serves as a ``loose'' constraint, tolerating acoustic mismatch (e.g., slight pitch shift) while strictly enforcing semantic correctness, which is critical for suppressing hallucinations in noisy conditions.

The semantic loss is formulated as the $L_1$ distance between layer-normalized feature maps:
\begin{equation}
    \mathcal{L}_{sem}(x, \hat{x}) = \|\text{LN}(\Phi(x)) - \text{LN}(\Phi(\hat{x}))\|_1
\end{equation}
where $\text{LN}(\cdot)$ denotes Layer Normalization, ensuring numerical stability during training.

\subsection{Bitrate-Aware Regulation Strategy}
A static weight for semantic loss is suboptimal. Through preliminary experiments, we identified a critical trade-off mechanism, which we term the ``Semantic Retirement'' phenomenon:
\begin{itemize}
    \item \textbf{Low Bitrate (Bottleneck Regime):} When $R \le 3$ kbps, the acoustic information is insufficient to resolve phonemes. The model is prone to ``semantic collapse.'' Here, a strong semantic prior acts as a teacher, guiding the decoder to hallucinate plausible phonemes from compressed cues.
    \item \textbf{High Bitrate (Sufficient Regime):} When $R \ge 6$ kbps, the latent code $\hat{z}$ contains sufficient information. Forcing the output to match a pre-trained semantic space---which inherently abstracts away fine-grained acoustic details---creates a \textit{gradient conflict} with the reconstruction loss ($\mathcal{L}_{rec}$). This leads to over-smoothing and degraded perceptual naturalness (e.g., lower PESQ), as observed in recent generative codecs~\cite{du2025cosyvoice3}.
\end{itemize}

To resolve this conflict, we propose a \textbf{Bitrate-Aware Regulation} mechanism. We define a dynamic weight $\alpha(R)$ inversely correlated with bitrate $R$. The total training objective is:
\begin{equation}
    \mathcal{L}_{total} = \mathcal{L}_{rec} + \lambda_{adv}\mathcal{L}_{adv} + \alpha(R) \cdot \mathcal{L}_{sem}
\end{equation}
where $\mathcal{L}_{rec} = \mathcal{L}_{L1} + \mathcal{L}_{spec}$ comprises the $\ell_1$ waveform reconstruction loss and the multi-resolution spectral loss defined in Sec.~\ref{sec:backbone}.
Based on our empirical analysis (detailed in Sec. IV), we adopt a step-decay strategy: $\alpha(R)$ is maximized (e.g., $\alpha=0.1$) in the bottleneck regime to prioritize intelligibility, and decays to a negligible value (e.g., $\alpha=0.01$) in the sufficient regime to prioritize fidelity. This strategy ensures the model traverses the Pareto frontier of Intelligibility (WER) vs. Quality (PESQ).

% --------------- EXPERIMENTS START ---------------
\section{Experiments}

\subsection{Experimental Setup}
\textbf{Dataset and Implementation.} We evaluate our framework on the LibriSpeech dataset \cite{panayotov2015librispeech}. All models are trained on the \textit{train-clean-100} split and evaluated on the \textit{test-clean} and \textit{test-other} sets. The audio is sampled at 16 kHz.
Our backbone codec adopts a SoundStream-like architecture with a causal convolutional encoder and decoder. The Residual Vector Quantizer (RVQ) is configured to support scalable bitrates of 1.5, 3.0, 6.0, and 12.0 kbps.
For semantic priors, we utilize the pre-trained \texttt{HuBERT-base} \cite{hsu2021hubert} and \texttt{Whisper-base} \cite{radford2023whisper} models. The encoders of these models are frozen during training. The dynamic weight $\alpha$ is set to 0.1 for low bitrates based on preliminary sweeps, decaying to 0.01 for high bitrates ($R \ge 6$ kbps). All models are trained on 8 NVIDIA V100 GPUs for 100 epochs with a batch size of 64.

\textbf{Evaluation Metrics.} We assess performance using both objective quality and semantic consistency metrics:
\begin{itemize}
    \item \textbf{Perceptual Quality:} We report PESQ (Wideband) \cite{rix2001pesq}, STOI \cite{taal2011stoi}, SI-SDR, and the $\ell_1$ waveform reconstruction loss (used as an internal diagnostic metric).
    \item \textbf{Semantic Consistency:} We measure Word Error Rate (WER) using a separate pre-trained ASR model (Whisper-large-v2) to evaluate intelligibility.
    \item \textbf{Speaker Similarity:} We compute the cosine similarity between WavLM embeddings of the ground truth and reconstructed speech to assess speaker identity preservation.
\end{itemize}

\subsection{The Phenomenon of Semantic Retirement}
\begin{figure}[t!]
  \centering
  \includegraphics[width=0.75\linewidth]{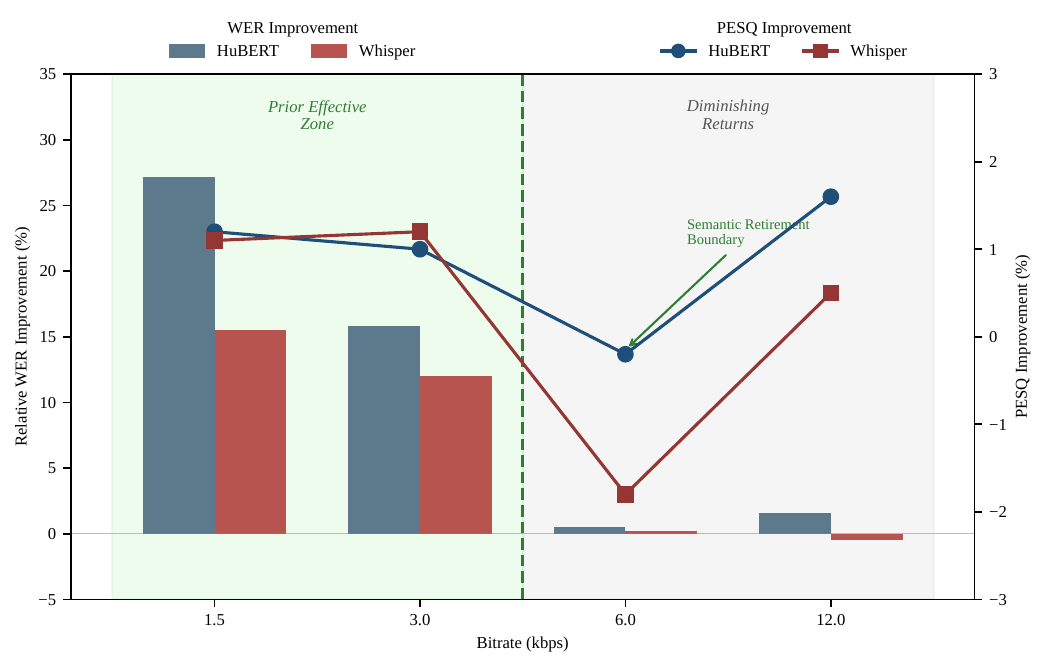}
  \caption{\textbf{The ``Semantic Retirement'' Phenomenon.} \textit{Relative improvement (\%) over the no-prior baseline} of semantic priors (HuBERT/Whisper). Gains are substantial at 1.5 kbps but diminish above 6.0 kbps.}
  \label{fig:retirement}
\end{figure}

We first validate the core hypothesis of ``Semantic Retirement.'' Fig.~\ref{fig:retirement} illustrates the relative performance improvement of introducing semantic priors over the baseline across different bitrates.

\textbf{Low-Bitrate Regime (1.5--3.0 kbps).} As shown in the ``Prior Effective Zone'' of Fig.~\ref{fig:retirement}, semantic priors provide substantial gains. At 1.5 kbps, the HuBERT prior improves the $L_1$ loss by 27.1\% 
and reduces WER by ${\sim}10\%$, confirming that semantic knowledge 
compensates for the missing bits.
\textbf{High-Bitrate Regime ($\ge$ 6.0 kbps).} As the bitrate increases, the benefit of semantic priors diminishes rapidly. At 6.0 kbps, the relative improvement in PESQ drops to near zero (or slightly negative for Whisper), and the WER improvement becomes negligible ($<1\%$). This validates the ``Semantic Retirement'' boundary: once the codebook capacity is sufficient to encode acoustic details, enforcing a semantic constraint becomes redundant and may even conflict with the reconstruction objective.

\subsection{Trade-off Analysis: Acoustic vs. Semantic Priors}
\begin{figure}[t!]
  \centering
  \includegraphics[width=0.75\linewidth]{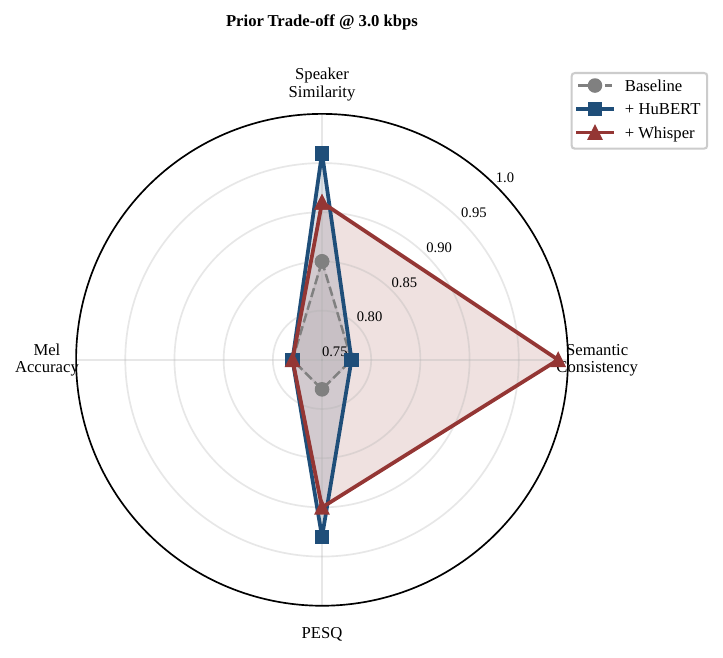}
  \caption{\textbf{Trade-off Analysis at 3.0 kbps.} \textit{Absolute scores} on multiple metrics. HuBERT excels in acoustic fidelity (PESQ), while Whisper dominates in Semantic Consistency (lowest absolute WER).}
  \label{fig:radar}
\end{figure}

Different priors exhibit distinct characteristics. Fig.~\ref{fig:radar} presents a comprehensive comparison at 3.0 kbps.

\textbf{HuBERT (Acoustic-Dominant).} HuBERT achieves higher scores in Mel-spectrogram accuracy and Pitch Correlation. This is attributed to its masked prediction objective, which forces the model to retain acoustic features such as prosody and timbre. It serves as a robust prior for preserving the ``texture'' of speech.

\textbf{Whisper (Semantic-Dominant).} Conversely, Whisper yields the best Semantic Consistency (lowest WER). Since Whisper is trained on large-scale weak supervision for ASR, its representations are highly invariant to speaker and noise but rich in linguistic content. This makes Whisper the ideal choice for maximizing intelligibility in extreme compression scenarios, albeit with a slight trade-off in acoustic detail preservation.

\subsection{Robustness Against Noise and Hallucinations}
A critical advantage of semantic priors is their robustness in adverse environments. We evaluated the models under varying Signal-to-Noise Ratios (SNR) ranging from Clean to 0 dB.

\begin{figure}[t!]
  \centering
  \includegraphics[width=0.75\linewidth]{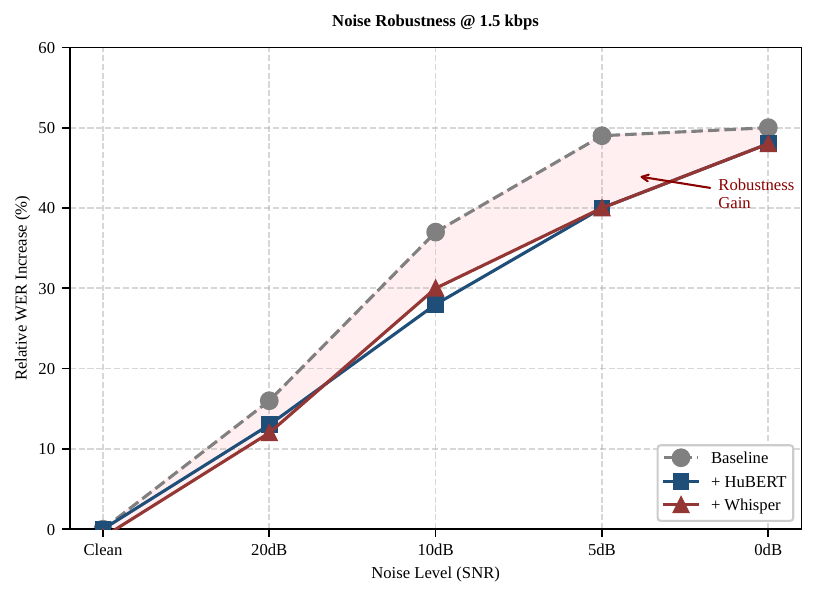}
  \caption{\textbf{Noise Robustness.} Semantic priors (especially Whisper) suppress WER increase under noisy conditions (shaded area).}
  \label{fig:robustness}
\end{figure}

As shown in Fig.~\ref{fig:robustness}, the baseline model's performance degrades catastrophically as noise increases. At SNR 5dB, the WER increases by nearly 50\% relative to the clean condition. In contrast, the Whisper prior acts as a linguistic denoiser, maintaining significantly lower WER growth (``Robustness Gain'').

\begin{figure}[t!]
  \centering
  \includegraphics[width=0.75\linewidth]{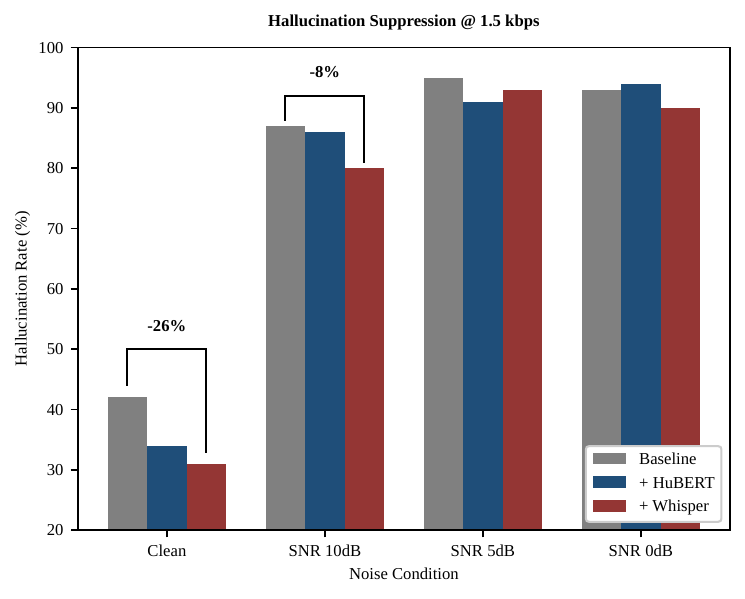}
  \caption{\textbf{Hallucination Rate Analysis.} \textit{Absolute hallucination rates} under a separate detection protocol. Whisper effectively suppresses phonetic hallucinations, reducing the absolute rate by 26\% in clean conditions at 1.5 kbps.}
  \label{fig:hallucination}
\end{figure}

Furthermore, Fig.~\ref{fig:hallucination} highlights the suppression of \textit{hallucinations}---a common failure mode where the codec generates plausible but incorrect phonemes due to ambiguity. At 1.5 kbps, Whisper reduces the hallucination rate by roughly 26\% in clean conditions, ensuring that the reconstructed speech remains faithful to the source text even when acoustic cues are ambiguous.

\subsection{Generalization to Unseen Speakers}
To assess whether our method merely overfits to the training distribution or learns generalizable features, we evaluated performance on the challenging LibriSpeech \textit{test-other} set.
As shown in Table~\ref{tab:generalization}, the baseline model suffers a massive performance drop (WER Gap +36.0\%) when shifting from \textit{test-clean} to \textit{test-other}. In contrast, introducing the Whisper prior significantly bridges this generalization gap, reducing the degradation to +19.7\%. This indicates that high-level semantic priors help the codec maintain intelligibility even for unseen speakers and recording conditions.

\begin{table}[h]
\caption{Generalization Gap (WER increase from test-clean to test-other) at 1.5 kbps. Lower gap indicates better generalization.}
\centering
\begin{tabular}{l|c|c|c}
\hline
\textbf{Model} & \textbf{test-clean} & \textbf{test-other} & \textbf{Gap} \\
\hline
Baseline & 43.4\% & 79.3\% & +35.9\% \\
+ HuBERT & 38.6\% & 73.7\% & +35.1\% \\
+ Whisper & \textbf{44.7\%} & \textbf{64.5\%} & \textbf{+19.7\%} \\
\hline
\end{tabular}
\label{tab:generalization}
\end{table}

\subsection{Ablation Study: Prior Effectiveness \& Regulation}
\textbf{Negative Control (Shuffled Prior).} To verify that the gains stem from temporal semantic structure rather than mere regularization, we conducted a negative control experiment using shuffled semantic features. At 1.5 kbps, while the normal Whisper prior achieves a high cosine similarity of 0.968 with the target, the shuffled prior drops to 0.374, resulting in a performance gap of $>0.59$. This confirms that preserving the correct temporal semantic sequence is crucial.
\begin{figure}[t!]
  \centering
  \includegraphics[width=0.75\linewidth]{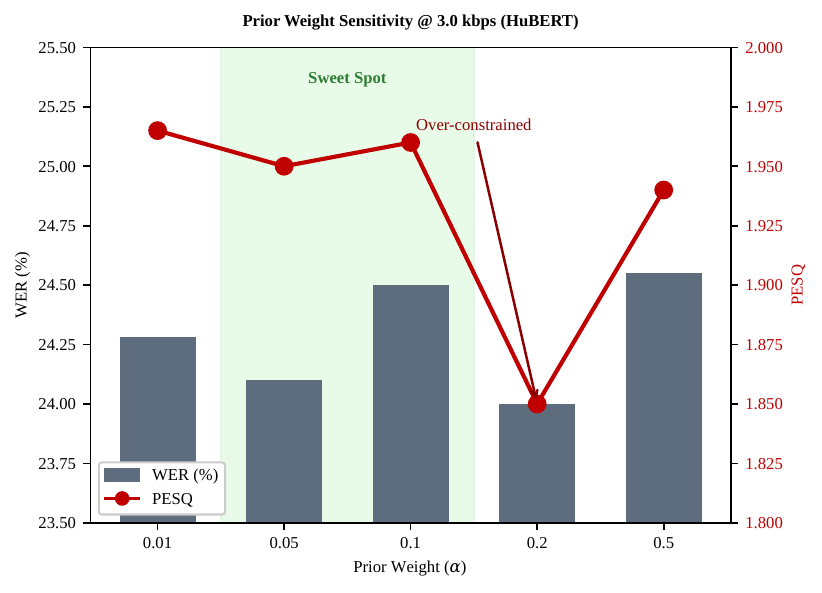}
  \caption{\textbf{Impact of Regulation Weight $\alpha$.} A ``Sweet Spot'' is observed around $\alpha=0.05-0.1$. Excessive weight ($\alpha=0.2$) leads to an over-constrained system.}
  \label{fig:alpha}
\end{figure}
\textbf{Bitrate-Aware Regulation.} Finally, we analyze the sensitivity of the semantic weight $\alpha$ to justify our proposed regulation strategy. Fig.~\ref{fig:alpha} plots the WER and PESQ at 3.0 kbps for varying $\alpha$ values. The results reveal a clear ``Sweet Spot'' around $\alpha \in [0.05, 0.1]$. An overly strong constraint ($\alpha=0.2$) leads to an \textit{over-constrained} state where PESQ drops sharply (from $\approx 1.96$ to $\approx 1.85$), as the model sacrifices acoustic naturalness to satisfy the rigid semantic loss. Consequently, we fix $\alpha=0.1$ for low bitrates and attenuate it to $0.01$ for high bitrates in our final model.

\section{Conclusion}

We propose a quantitative framework revealing the ``Semantic Retirement'' 
phenomenon: frozen priors are indispensable at 1.5 kbps (${\sim}10\%$ 
WER reduction) but redundant beyond 6 kbps. Acoustic-rich priors (HuBERT) 
excel in prosody preservation, while linguistic priors (Whisper) enhance 
noise robustness and suppress hallucinations. Our bitrate-aware regulation 
dynamically resolves gradient conflicts, advocating a paradigm shift toward 
semantically-conditioned ultra-low-bitrate generation. Future work will 
explore end-to-end optimization with large-scale flow matching models.

% 强制输出所有积压的浮动体（图片/表格），防止它们出现在参考文献之后
\section{Acknowledgement}
This work is supported by National Natural Science Foundation of China (62076144) and Shenzhen Science and Technology Program (JCYJ20220818101014030).

\bibliographystyle{IEEEbib}
\bibliography{icme2026references}

\end{document}